\documentclass[%
prd,
endfloats*,
showpacs,
showkeys,
preprint,
nofootinbib 
] {revtex4}
\usepackage{graphicx,hyperref,amssymb,amsmath,enumerate}

\begin{document}

\title{Sterile neutrinos in neutrinoless double beta decay}

\date{\today}

\author{P. Bene\v s}
\email{Petr.Benes@utef.cvut.cz} 
\altaffiliation{On  leave of absence from 
Institute of Experimental and Applied Physics, Czech Technical
University, Horsk\'a 3a/22, 128 00 Praha 2, Czech republic} 
\affiliation{Institute f\"ur Theoretische
Physik der Univesit\"at T\"ubingen, Auf der Morgenstelle 14,
D-72076 T\"ubingen, Germany}
\author{Amand Faessler}
\affiliation{Institute f\"ur Theoretische Physik der Univesit\"at
T\"ubingen, Auf der Morgenstelle 14, D-72076 T\"ubingen, Germany}
\author{S. Kovalenko}
\email{sergey.kovalenko@usm.cl} 
\affiliation{Departamento de F{\'
i}sica, Universidad T{\' e}cnica Federico Santa Mar{\' i}a,
Casilla 110-V, Valpara{\' i}so, Chile}
\author{F. \v Simkovic}
\email{fedor.simkovic@fmph.uniba.sk}
\altaffiliation{On  leave of absence from Department of Nuclear
Physics, Comenius University, Mlynsk\'a dolina F1, SK--842 15
Bratislava, Slovakia} \affiliation{Institute f\"ur Theoretische
Physik der Univesit\"at T\"ubingen, Auf der Morgenstelle 14,
D-72076 T\"ubingen, Germany}

\begin{abstract}
We study possible contribution of the Majorana neutrino mass eigenstate $\nu_h$ dominated 
by a sterile neutrino component to neutrinoless double beta ($0\nu\beta\beta$) decay. 
From the current experimental lower bound on the $0\nu\beta\beta$-decay half-life of $^{76}$Ge we derive
stringent constraints on the $\nu_h-\nu_e$ mixing in a wide region of the values 
of $\nu_h$ mass. We discuss cosmological and astrophysical status of $\nu_h$ 
in this mass region. 
 \end{abstract}



\maketitle


The sterile right-handed neutrinos $\nu_R$, not participating in the electroweak interactions, are natural
candidates for the extension of the standard model (SM) field contents. They are very typical for  
various scenarios of the physics beyond the SM. It has long been recognized 
that the sterile neutrinos may have plenty of phenomenological implications. 
One of the celebrated examples is the seesaw generation of tiny Majorana neutrino masses for the three 
observable very light neutrinos via a very large Majorana mass term of the right-handed neutrinos. 
In this case together with the very light neutrinos there also appear very heavy Majorana 
neutrino mass states with the typical masses of $\sim 10^{12}$ GeV. However, in a more general case of 
the sterile-active neutrino mixing there may also appear 
additional neutrino states $\nu_h$ with arbitrary masses $m_{h}$. The neutrino mass eigenstates $\nu_h$,
dominated by the sterile $\nu_R$ neutrino weak eigenstates, contain   
some admixture of the active $\nu_{e,\mu,\tau}$ neutrino species that allows the $\nu_h$ to contribute to various processes, 
in particular, to those which are forbidden in the SM  by the Lepton Number or Lepton Flavor Violation (LNV or LFV). 
They may also modify the interpretation of cosmological and astrophysical observations. Therefore, the masses $m_{h}$
of $\nu_h$ neutrino states and their mixing with the active neutrinos are subject to various experimental 
as well as cosmological and astrophysical constraints. 
%
%
Various implications of the sterile neutrinos have been 
extensively studied in the literature within different scenarios of neutrino mass generation \cite{sterile-th,Bamert:1994qh}. 

Experimentally, the $\nu_h$ neutrino states 
can be searched for as peaks in differential rates of various processes and by direct production of $\nu_h$ 
followed by their decays in detector (for summary see Ref. \cite{PDG}).
The $\nu_h$ can also give rise to a significant enhancement of the total rates of certain 
processes if their masses, $m_{h}$, have the values in appropriate regions specific for each of these processes 
\cite{SP,SLGFK}. 
This effect would be especially pronounced in reactions that are forbidden in the SM. 
The LNV and/or LFV processes belong to this category. Many of them are 
stringently restricted by experiment and allow one to derive stringent limits on the $\nu_h$ contribution.
Unprecedented sensitivities have been reached in the experiments searching for neutrinoless nuclear 
double beta decay ($0\nu\beta\beta$-decay) \cite{klaepja12}. The current limits on the half-life of $0\nu\beta\beta$-decay 
are known to stringently constrain various parameters of the LNV physics, in particular, those which correspond 
to the Majorana neutrino contributions.

The sterile neutrino contribution to $0\nu\beta\beta$-decay has been first studied in the pioneer work Ref. \cite{Bamert:1994qh}
within a specific model with the two sterile neutrinos mixed only with the $\nu_e$ flavor state. 
In the present paper we develop this analysis by a more careful treatment of the nuclear $0\nu\beta\beta$-decay 
matrix elements on the basis of a well motivated nuclear structure approach known as 
the renormalized proton-neutron Quasiparticle Random Phase Approximation(QRPA) \cite{simprc60}. Also, we adopt a general 
phenomenological approach to the neutrino mass generation without referring to a specific model.
We consider an extension of the SM with the three left-handed weak
doublet neutrinos $\nu'_{Li} = (\nu'_{Le},\nu'_{L\mu},\nu'_{L\tau})$
mixed with $n$ species of the SM singlet right-handed neutrinos
$\nu'_{Ri}=(\nu'_{R1},...\nu'_{Rn})$. In general, in this scenario there appear $n+3$ Majorana 
neutrino mass eigenstates $\nu_i = U^*_{ki}\nu'_k$, where $U_{ki}$ is the corresponding mixing matrix.  
%
%
For simplicity, we assume that in addition to the three conventional light neutrinos there exists only one 
Majorana neutrino mass eigenstate $\nu_h$, dominated by the sterile neutrino species, with an arbitrary mass $m_{h}$, 
which may mix with all the active neutrino weak eigenstates, $\nu_{e,\mu,\tau}$. We study possible contribution of 
this $\nu_h$ neutrino
state to $0\nu\beta\beta$-decay via a nonzero admixture of $\nu_e$ weak eigenstate. From the non-observation of 
this LNV process we derive stringent limits on the $\nu_h-\nu_{e}$ mixing matrix element $U_{e h}$ 
in a wide region of the values of $m_{h}$.
   


%
The contribution of $\nu_h$ neutrino state to the $0\nu\beta\beta$-decay amplitude is described 
by the standard Majorana neutrino exchange between the two
$\beta$-decaying neutrons. The corresponding half-life formula for a transition to 
the ground state of final nucleus is given by the expression
\begin{equation}
[T_{1/2}^{0\nu}]^{-1} = G_{01} 
\left|U_{eh}^2 \frac{m_h}{m_e} M^{0\nu}(m_h)\right|^2.
\label{eq:1}   
\end{equation}
Here, $m_e$ is the mass of electron and
$G_{01}$ is the integrated kinematical factor (for its 
numerical values see e.g. Ref. \cite{simprc60}). 
The nuclear matrix element $M^{0\nu}(m_h)$ we write in the form
\begin{equation}
M^{0\nu} (m_h) = \left\langle 
H_{F} (m_h, r_{12})  {\bf 1} + 
H_{GT} (m_h, r_{12}) {\bf \sigma}_{12} +
H_{T} (m_h, r_{12}) {\bf S}_{12}
\right\rangle
\label{eq:5}   
\end{equation}
with the three terms corresponding to the Fermi, Gamow-Teller, and tensor contributions, which depend on
the neutrino mass $m_{h}$ and on the relative coordinates defined as
\mbox{
${\bf r}_{12} = {\bf r}_1-{\bf r}_2,\  
r_{12} = |{\bf r}_{12}|, \ 
\hat{{\bf r}}_{12} = {\bf r}_{12}/r_{12},
$}
where ${\bf r}_1$ and ${\bf r}_2$ are the coordinates of beta decaying nucleons.
We also introduced the operators: 
$
S_{12} = 3({\vec{ \sigma}}_1\cdot \hat{{\bf r}}_{12})
       ({\vec{\sigma}}_2 \cdot \hat{{\bf r}}_{12})
      - \sigma_{12}, ~~~ \sigma_{12}=
{\vec{ \sigma}}_1\cdot {\vec{ \sigma}}_2.
$

The neutrino-exchange potentials $H_{K}(m_h,r_{12})$, with $K = F, GT, T$, have the following form:
\begin{eqnarray}   
H_{K}(m_h, r_{12})=\frac{2}{\pi g_A^2}
{R} \int_{0}^{\infty} 
f_K(qr_{12})
\frac{\ h_{K}(q^2)~q^2} 
{\sqrt{q^2+m^2_h} (\sqrt{q^2+m^2_h}
+E^m_{J} - (E^i_{g.s.} + E^f_{g.s.})/2)}
~d{q}. ~~
\label{eq:7}   
\end{eqnarray}
where $f_K(qr_{12}) = j_0(qr_{12})$, $j_0(qr_{12})$ and $j_2(qr_{12})$ for $K=F$, $GT$ and $T$,
respectively ($j_0$ and $j_2$ are spherical bessel functions). $E^i_{g.s.}$, $E^f_{g.s.}$, and 
$E^{m}_{J}$ are the energies of initial, final, and intermediate nuclear states, respectively. 
The mean nuclear radius is given by $R = r_0 A^{1/3}$ 
with $r_0 = 1.1~ fm$, and $g_A=1.25$ is the axial nucleon constant. 
The functions $h_{K}(q^2)$ are 
\begin{eqnarray}
h_{F}({ q}^{~2} ) &=& - g^2_{V} ({q}^{~2}),~~
h_{T}({ q}^{~2} ) =
 \frac{2}{3}\frac{g_A ({ q}^{~2}) g_P ({ q}^{~2}) { q}^{~2}}{2 m_p}
- \frac{1}{3}\frac{g^2_P ({ q}^{~2}) { q}^{~4}}{4 m^2_p}
+ \frac{1}{3} \frac{g^2_M ({ q}^{~2}) { q}^{~2}}{4m^2_p}, \nonumber  \\
h_{GT}({ q}^{~2} ) &=& g^2_A ({q}^{~2}) -\frac{2}{3}\frac{g_A ({ q}^{~2}) 
g_P ({ q}^{~2}) { q}^{~2}}{2 m_p} +
\frac{1}{3}\frac{g^2_P ({ q}^{~2}) {q}^{~4}}{4 m^2_p}+
\frac{2}{3} \frac{g^2_M ({ q}^{~2}) { q}^{~2} }{4m^2_p}.
\label{eq:8}   
\end{eqnarray}
For the nucleon form factors we use the standard parameterizations: \\
$
g_V({ q}^{~2}) = (1+{ q}^{~2}/{\Lambda^2_V})^{-2},\ 
g_A({ q}^{~2}) = g_A(1+{ q}^{~2}/{\Lambda^2_A})^{-2}, \ 
g_M({ q}^{~2}) = (\mu_p-\mu_n) g_V({ q}^{~2}), \\  
g_P({ q}^{~2}) = 2 m_p g_A({ q}^{~2})({ q}^{~2} + m^2_\pi)^{-1},
$
where 
$(\mu_p-\mu_n) = 3.70$, $\Lambda_V = 0.84$ GeV, $\Lambda_A = 1.09$ GeV,
and $m_\pi$ is the pion mass. 

We calculated the nuclear matrix element in Eq.~(\ref{eq:5}) for the transition
$^{76}$Ge$(0^+_{g.s.}) \rightarrow ^{76}$Se$(0^+_{g.s.})$ within the proton-neutron
renormalized QRPA \cite{simprc60}. We analyzed the single-particle model space which
involves the full $2-4\hbar\omega$ major shells both for protons and neutrons.
The single particle energies were obtained using the Coulomb--corrected 
Woods--Saxon potential. We calculated the two-body G-matrix elements from the Bonn
one-boson exchange potential on the basis of the Brueckner theory.
Since the considered model space is finite the pairing
interactions have been adjusted to fit the empirical pairing
gaps. In addition, we renormalized the particle-particle and particle-hole channels of the G-matrix
interaction of the nuclear Hamiltonian following the
procedure outlined in Ref. \cite{rodprc68}. In this way we obtained the nuclear matrix element with 
a weak dependence on the nuclear structure parameter space.
Finally, in order to take into account the two-nucleon correlations we 
multiplied the transition operators by the square of the correlation 
Jastrow-like function \cite{jastr}. 

\begin{figure}[t]
  \begin{center}
    \includegraphics[height = 12.0cm]{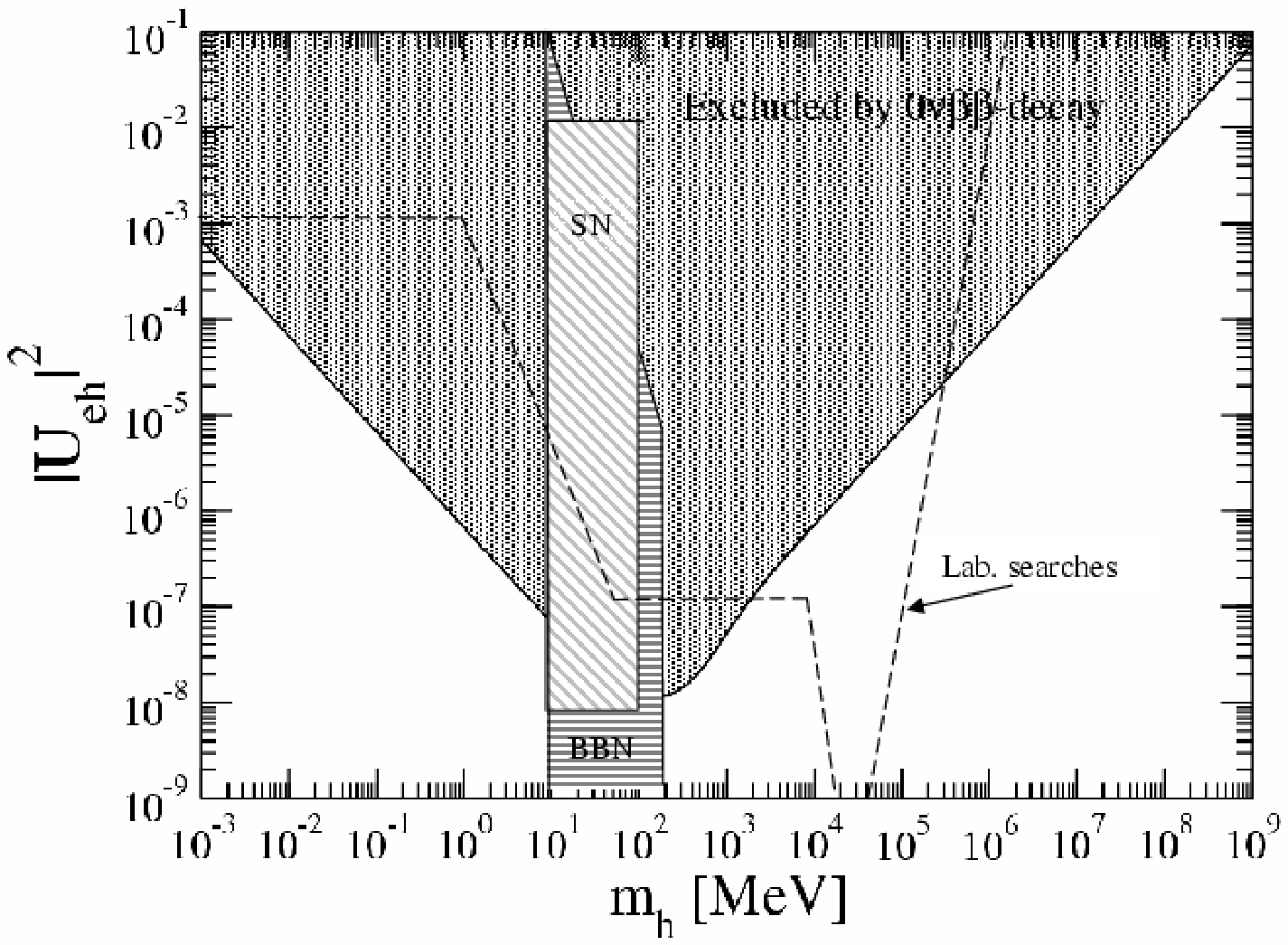}
  \end{center}
  \caption{Exclusion plots in the $|U_{eh}|^2-m_h$ plane. 
The shaded regions are excluded by the $0\nu\beta\beta$-decay [present result], and by 
the big bang nucleosynthesis and the SN 1987A neutrino observations. The dashed 
line mark typical upper limits from the laboratory searches for massive neutrinos.
\label{fig.2}}
\end{figure}

Having the nuclear matrix element $M^{0\nu}(m_h)$ calculated, we can derive 
the $0\nu\beta\beta$ decay limits on the mass $m_h$ of the  $\nu_h$ neutrino 
and its mixing $U_{e h}$ with the $\nu_e$ neutrino weak eigenstate.
Applying the presently best lower bound on the $0\nu\beta\beta$-decay half-life of $^{76}$Ge  \cite{klaepja12} 
\begin{eqnarray}\label{HD-M}
T_{1/2}^{0\nu}\geq T_{1/2}^{0\nu-exp} = 1.9\times 10^{25} \mbox{years}
\end{eqnarray}
we derive from Eq. (\ref{eq:1}) the $|U_{e h}|^2-m_h$ exclusion plot shown in Fig.~\ref{fig.2}. 
In Fig.~\ref{fig.2}
we also show typical domains excluded by some other experiments summarized in 
Ref.~\cite{PDG}. However, these domains are just indicative, because most of the previous bounds were obtained for 
some fixed values of $m_h$. For convenience, we interpolated this set of experimental points by continuous curves
in different intervals of $m_h$. 
Let us also point out the following. The constraints listed in Ref. \cite{PDG} are based on the 
searches for peaks in differential rates of various processes and the direct production of $\nu_h$ states 
followed by their decays in a detector. In Ref. \cite{SP} it was argued that in 
this case the results of data analysis depend on the $\nu_h$
total decay width, including the neutral current decay channels. The latter have not been properly 
taken into account in the derivation
of the mentioned experimental constraints. 
However, the neutral current $\nu_h$ decay channels introduce the dependence of the final results on all of 
the mixing matrix elements $U_{e h}, U_{\mu h}$ and $U_{\tau h}$. In this situation one cannot extract individual 
limits for these matrix elements without some additional assumptions, introducing a significant uncertainty. 
In contrast, our $0\nu\beta\beta$-decay 
limits involve only $U_{e h}$ mixing matrix element and, therefore, are free of the mentioned uncertainty. 
This is because in $0\nu\beta\beta$-decay intermediate Majorana neutrinos are always off-mass-shell states and their decay 
widths are irrelevant.  

It is well known that massive neutrinos may have
important cosmological and astrophysical implications.
They are expected to contribute to the mass density of
the universe, participate in cosmic structure formation,
big-bang nucleosynthesis, supernova explosions, imprint themselves
in the cosmic microwave background etc.\ (see, for instance,
Refs.  \cite{Raffelt,Dolgov,afp:2001} and references therein). This implies certain constrains on the neutrino masses
and mixing, which, however, for unstable neutrinos, such as $\nu_h$ studied in the present paper, are 
quite weak and involve various uncertainties.
The Big-bang nucleosynthesis(BBN) and the SN 1987A neutrino signal 
provide the most stringent and certain constraints \cite{Dolgov} for this case.
The corresponding constraints on $\nu_h$ parameters have been derived
in Ref. \cite{Dolgov} for the restricted mass regions:  $10 \mbox{MeV}\leq m_h\leq 200$ MeV 
for the BBN 
and $10 \mbox{MeV}\leq m_h\leq 100$ MeV for the SN 1987A cases. We present the corresponding exclusion plots
in Fig.~\ref{fig.2} for comparison with our limits. 

As seen from Fig.~\ref{fig.2}, our $0\nu\beta\beta$-decay limits cover much wider region of $\nu_h$ masses $m_h$
and exclude the parts of the $|U_{e h}|^2-m_h$ parameter space previously unconstrained by the laboratory 
experiments, and the cosmological and astrophysical observations. 


In summary, we examined the possible contribution of the sterile neutrino dominated Majorana mass eigenstate $\nu_h$ to
the $0\nu\beta\beta$ decay. 
From the currently most stringent lower bound on the $0\nu\beta\beta$-decay half-life of $^{76}$Ge we derived upper limits on 
the $\nu_h-\nu_e$ neutrino mixing matrix element $|U_{e h}|^2$ in a wide region of the values of $m_h$, excluding 
previously unconstrained domains in the $|U_{e h}|^2-m_h$ plane. Our constraints are free of the uncertainties, 
related to the assumptions on the values of the other neutrino mixing matrix elements $U_{\mu h}$ and $U_{\tau h}$, 
typical for the other laboratory constraints of this type.

\begin{acknowledgments}
This work was supported in part by Fondecyt (Chile) under
grant No.~1030244,  by the DFG (Germany) under contract 436 SLK 113/8,
by the VEGA Grant agency (Slovakia) under contract No.~1/0249/03
and by the Grant Agency of the Czech Republic Grant No.
202/02/0157.

\end{acknowledgments}

\begin{thebibliography}{qq}
%
\bibitem{sterile-th} D.O. Caldwell and R.N. Mohapatra, Phys. Rev. D {\bf 48}, 3259 (1993); J. T. Peltoniemi and J. 
W. F. Valle, Nucl. Phys. B {\bf 406}, 409 (1993); J. T. Peltoniemi, D. Tommasini and J. W. F. 
Valle, Phys. Lett. B {\bf 298}, 383 (1993);
M. Sorel, J. M. Conrad and M. Shaevitz, Phys.Rev. D {\bf 70}, 073004 (2004); 
R. N. Mohapatra and S. Nasri, arXiv:hep-ph/0407194;
K. S. Babu and G. Seidl, Phys.Rev. D {\bf 70}, 113014  (2004); 
Phys. Lett. B {\bf 591}, 127 (2004); 
K. S. Babu and R. N. Mohapatra, Phys. Lett. B {\bf 522}, 287 (2001); 
S. Godfrey and S. H. Zhu, arXiv:hep-ph/0405006; 
W. Krolikowski, Acta Phys. Polon. B {\bf 35}, 1675 (2004),
ibid 35, 2241 (2004);
K. L. McDonald, B. H. J. McKellar and A. Mastrano, Phys.Rev. D {\bf 70}, 053012 (2004);
V. Barger, D. Marfatia and K. Whisnant, Phys. Lett. B {\bf 576}, 303 (2003); 
G. J. Stephenson, T. Goldman, B. H. J. McKellar and M. Garbutt, arXiv:hep-ph/0307245; 
H. Paes, L. G. Song and T. J. Weiler, Phys. Rev. D {\bf 67}, 073019 (2003);
A. Donini and D. Meloni, Nucl. Instrum. Meth. A {\bf 503}, 230 (2001);
Eur. Phys. J. C {\bf 22}, 179 (2001);
M. Frank, I. Turan and M. Sher, arXiv:hep-ph/0412090. 
%
%
\bibitem{Bamert:1994qh}
P.~Bamert, C.~P.~Burgess and R.~N.~Mohapatra,
Nucl.\ Phys.\ B {\bf 438}, 3 (1995)
%
\bibitem{PDG} Review of Particle Physics, Phys. Lett. B {\bf 592}, 1 (2004).  
%
%
%
\bibitem{SP} C. Dib, V. Gribanov, S. Kovalenko, and I. Schmidt,
  Phys. Lett. B {\bf 493}, 82 (2000); V. Gribanov, S. Kovalenko, and I. Schmidt,
  Nucl. Phys. B {\bf 607}, 355 (2001).
\bibitem{SLGFK}   F. \v Simkovic, V.E. Lyubovitskij, Th. Gutsche, A. Faessler,
  and S. Kovalenko, Phys. Lett. B {\bf 544}, 121 (2002).
%
%
\bibitem{klaepja12} Heidelberg-Moscow Collaboration, H.V. Klapdor-Kleingrothaus,
 et al., Eur. Phys. J. A {\bf 12}, 147 (2001). 
%
%
\bibitem{simprc60} F. \v Simkovic, G. Pantis, J.D. Vergados, and A. Faessler,
  Phys. Rev. C {\bf 60}, 055502 (1999).
%
%
%
%
%
%
%
\bibitem{rodprc68} V.A. Rodin, A. Faessler, F. \v Simkovic, and P. Vogel,
  Phys. Rev. C {\bf 68}, 044302 (2003).
%
%

\bibitem{jastr} G.A.Miller and J.E. Spencer, Ann. Phys.{\bf 100}, 562 (1976).

\bibitem{bilenky:2004} S. M. Bilenky, A. Faessler, F. \v Simkovic,
  Phys. Rev. D {\bf 70}, 033003 (2004).

\bibitem{Raffelt}  G.G. Raffelt,
{\it Stars as Laboratories for Fundamental Physics}, University
of Chicago Press, 1996;
Ann. Review of Nuclear and Particle Science, { 49} (1999).
%
\bibitem{Dolgov} A.D. Dolgov, S.H. Hansen, G. Raffelt, D.V. Semikoz,
Nucl.Phys. B {\bf 590},  562 (2000); Nucl.Phys. B {\bf 580}, 331  (2000) 
and references therein.

\bibitem{afp:2001} K. Abazajian, G. M. Fuller, and M. Patel, Phys.Rev. D {\bf 64}, 023501 (2001).

\end{thebibliography}
\end{document}